\documentclass[sigconf]{acmart} 

\usepackage{tikz}
\usetikzlibrary{decorations.pathmorphing} 
\usetikzlibrary{fit}					
\usetikzlibrary{backgrounds}	

\newcommand*\rotb{\rotatebox{90}}
\newcommand*\rot{\rotatebox{270}}
\newcommand*\rotnothing{\rotatebox{360}}

\definecolor{rel}{RGB}{204, 0, 0}
\definecolor{val1}{RGB}{0,90,191}
\definecolor{val2}{RGB}{204, 0, 102}

\AtBeginDocument{%
  \providecommand\BibTeX{{%
    \normalfont B\kern-0.5em{\scshape i\kern-0.25em b}\kern-0.8em\TeX}}}

\copyrightyear{2023}
\acmYear{2023}
\setcopyright{rightsretained}
\acmConference[DLfM 2023]{10th International Conference on Digital Libraries for Musicology}{November 10, 2023}{Milan, Italy}
\acmBooktitle{10th International Conference on Digital Libraries for Musicology (DLfM 2023), November 10, 2023, Milan, Italy}\acmDOI{10.1145/3625135.3625141}
\acmISBN{979-8-4007-0833-6/23/11}

\acmPrice{15.00}
\acmISBN{978-1-4503-XXXX-X/18/06}

\begin{document}

\title{Sounding Out Reconstruction Error-Based Evaluation of Generative Models of Expressive Performance}

\author{Silvan David Peter}
\email{silvan.peter@jku.at}
\orcid{0009-0000-8328-291X}
\affiliation{
  \institution{Johannes Kepler University}
  \city{Linz}
  \country{Austria}
}

\author{Carlos Eduardo Cancino-Chac\'on}
\email{carlos_eduardo.cancino_chacon@jku.at}
\orcid{0000-0001-5770-7005}
\affiliation{
  \institution{Johannes Kepler University}
  \city{Linz}
  \country{Austria}
}

\author{Emmanouil Karystinaios}
\email{emmanouil.karystinaios@jku.at}
\orcid{0000-0001-9354-8953}
\affiliation{
  \institution{Johannes Kepler University}
  \city{Linz}
  \country{Austria}
}

\author{Gerhard Widmer}
\email{gerhard.widmer@jku.at}
\orcid{0000-0003-3531-1282}
\affiliation{
  \institution{Johannes Kepler University}
  \city{Linz}
  \country{Austria}
}

\renewcommand{\shortauthors}{Trovato and Tobin, et al.}

\begin{abstract}

Generative models of expressive piano performance are usually assessed by comparing their predictions to a reference human performance.
A generative algorithm is taken to be better than competing ones if it produces
performances that are closer to a human reference performance.
However, expert human performers can (and do) interpret music in different ways, making for different possible references, and quantitative closeness is not necessarily aligned with perceptual similarity, raising concerns about the validity of this evaluation approach.
In this work, we present a number of experiments that shed light on this problem.
Using precisely measured high-quality performances of classical piano music,
we carry out a listening test indicating that listeners can sometimes perceive subtle performance difference that go unnoticed under quantitative evaluation.
We further present tests that indicate that such evaluation frameworks show a lot of variability in reliability and validity across different reference performances and pieces. 
We discuss these results and their implications for quantitative evaluation, and hope to foster a critical appreciation of the uncertainties involved in quantitative assessments of such performances within the wider music information retrieval (MIR) community.

\end{abstract}



\maketitle

\section{Introduction}\label{sec:introduction}

The recent years have seen the creation and publication of several corpora of precisely measured and score-aligned piano performances within MIR and  digital musicology communities~\cite{Peter-2023, Jeongetal:2019,zhang2022atepp}.
This renewed interest in computational models of expressive piano performance, in particular the data-driven kind.
Yet it also rekindled concerns surrounding the direct applicability of large scale data processing and machine learning techniques to this type of data. 

This paper addresses one such concern, namely issues of quantitatively evaluating generative models of expressive piano performance (GMEPP) at scale.
Quantitative evaluation in itself is nothing new, GMEPPs are routinely evaluated in terms of how close their predictions are to actual human expert performances.
This closeness is generally estimated with figures of merit such as reconstruction errors~\cite{CancinoChacon:2017ht,Jeongetal:2019} or likelihood functions~\cite{Kim:2013wi,Grindlay:2006eea,Flossmann:2013ho}.

One issue with this type of evaluation arises from the fact that having a model produce a performance that is \emph{numerically close} (in some aspect yet to be clarified) to an expert piano performance --- i.e., a model that does well according to generally accepted figures of merit --- possibly misses the mark of GMEPP; the goal of producing convincing, musical, and consistent performances for human listeners. 
Evaluating by asking such listeners is, however, only an option in a minority of situations, and most of the time the training, development, and evaluation of GMEPP requires scalable, automated metrics.
This potential goal misalignment raises at least two problems:  
is a measured distance to a human reference performance related to the perceptual similarity of performances? 
And is the choice of an arbitrary human reference performance immaterial for evaluative outcome?

These questions tap into profound epistemic, perceptual, and axiological issues beyond the scope this article.
What we can and do address in the following, are three smaller, but nevertheless operatively useful questions about current reconstruction error-based evaluation (REE) techniques:

\begin{itemize}
\item Can listeners discern performances that are indistinguishable under REE?
\item To what extent does REE reliably favor [performances by] the same model under different reference and piece conditions?
\item To what extent does REE validly identify the [performances by] expert pianists under different reference and piece conditions?
\end{itemize}

To assess these issues, we set up two experiments.
First, a listening test asking participants to identify expert performances in pairs of expert and artificially generated performances.
Second, we investigate the reliability and validity of REE evaluation, using the previously assessed artificially generated performances as negatives.
We discuss the results of these experiments in the context of the literature on  and perception of expressive performance and we identify potential steps to improve quantitative evaluation of GMEPP.
With more, larger, and ecologically valid (i.e., stemming from realistic performance scenarios) datasets of expressive piano performance becoming publicly available and used by the wider MIR community, we hope this discussion to foster a critical appreciation of the uncertainties involved in quantitative assessments of such performances.

The rest of this paper is structured as follows:
Section \ref{sec:evaluation} details the framework of quantitative GMEPP evaluation as investigated in this article.
Section \ref{sec:tce} describes how we extract and preprocess expressive parameters from recordings of expressive expert performances.
Section \ref{sec:experiments1} describes the performance discernment listening test and
section \ref{sec:experiments2} details the reliability and validity experiments.
Finally, section \ref{sec:discussion} discusses these results for evaluation of GMEPP and concludes this article.
The audio files, code, and data is available at~\url{https://github.com/CPJKU/performance_similarity_dlfm23}.

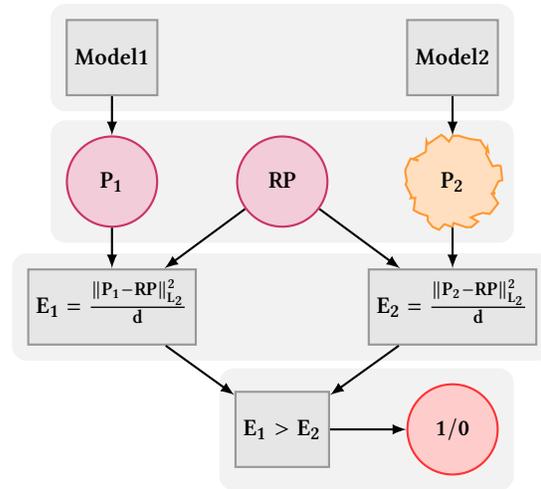
\begin{figure}[h]
\tikzstyle{decision}=[circle,
                                    thick,
                                    minimum size=1.2cm,
                                    draw=red!80,
                                    fill=red!20]

\tikzstyle{input}=[circle,
                                    thick,
                                    minimum size=1.2cm,
                                    draw=purple!80,
                                    fill=purple!20]

\tikzstyle{matrx}=[rectangle,
                                    thick,
                                    minimum size=1cm,
                                    draw=gray!80,
                                    fill=gray!20]

\tikzstyle{noise}=[circle,
                                    thick,
                                    minimum size=1.2cm,
                                    draw=orange!80,
                                    fill=orange!25,
                                    decorate,
                                    decoration={random steps,
                                                            segment length=2pt,
                                                            amplitude=2pt}]

\tikzstyle{background}=[rectangle,
                                                fill=gray!10,
                                                inner sep=0.2cm,
                                                rounded corners=2mm]

\begin{tikzpicture}[>=latex,text height=1.5ex,text depth=0.25ex, scale= 1.5]
  \matrix[row sep=0.5cm,column sep=0.5cm] {
        \node (mod1) [matrx] {$\mathbf{Model 1}$};       & &
        \node (mod2)   [matrx] {$\mathbf{Model 2}$};    & 
        \\
        \node (p1) [input]{$\mathbf{P_1}$}; &
        \node (rp)   [input]{$\mathbf{RP}$}; &
        \node (p2) [noise] {$\mathbf{P_2}$};&
        \\
        \node (e1) [matrx] {$\mathbf{E_1=\frac{\|P_1 - RP\|_{L_2}^{2}}{d}}$};       &&
        \node (e2)   [matrx] {$\mathbf{E_2 =\frac{\|P_2 - RP\|_{L_2}^{2}}{d}}$};      & 
          \\
          
        &
        \node (comp) [matrx] {$\mathbf{E_1 > E_2}$};       &\node (winner) [decision]{$\mathbf{1/0}$}; &
        \\
    };

    \path[->]
        (mod1) edge[thick] (p1)	
        (mod2) edge[thick] (p2)
        (p1) edge[thick] (e1)
        (rp) edge[thick] (e1)
        (p2) edge[thick] (e2)
        (rp) edge[thick] (e2)
        (e1) edge[thick] (comp)
        (e2) edge[thick] (comp)
        (comp) edge[thick] (winner)
        ;

    \begin{pgfonlayer}{background}
        \node [background,
                    fit=(mod1) (mod2)] {};
                \node [background,
                    fit=(p1) (p2)] {};
                  \node [background,
                    fit=(e1) (e2)] {};   
                  \node [background,
                    fit=(comp) (winner)] {};  
                    
    \end{pgfonlayer}

\end{tikzpicture}
\caption{
Schematic representation of our framework for two model evaluation.
These frameworks are commonly used for the comparison of two or more candidate models of expressive performance. In our experiments, however, the models are specifically designed for their known ground truth wrt evaluation (in the sense discussed in \ref{sec:evaluation}): Model 1 only produces expert performances (purple), model 2 only randomly sampled performances (orange), i.e. model 1 is the musically valid one. The two models produce a performance each ($P_1$ and $P_2$). The MSE of the performances with respect to an expert \emph{reference performance} ($RP$) is measured ($E_1$ and $E_2$, row 3). The comparison of error terms (row 4) outputs a Boolean decision value (red).
}

\label{fig:mod_eval}
\end{figure}

\section{A Framework of Quantitative Evaluation}\label{sec:evaluation}

\begin{figure*}[t]
\begin{center}
\includegraphics[width=0.75\linewidth]{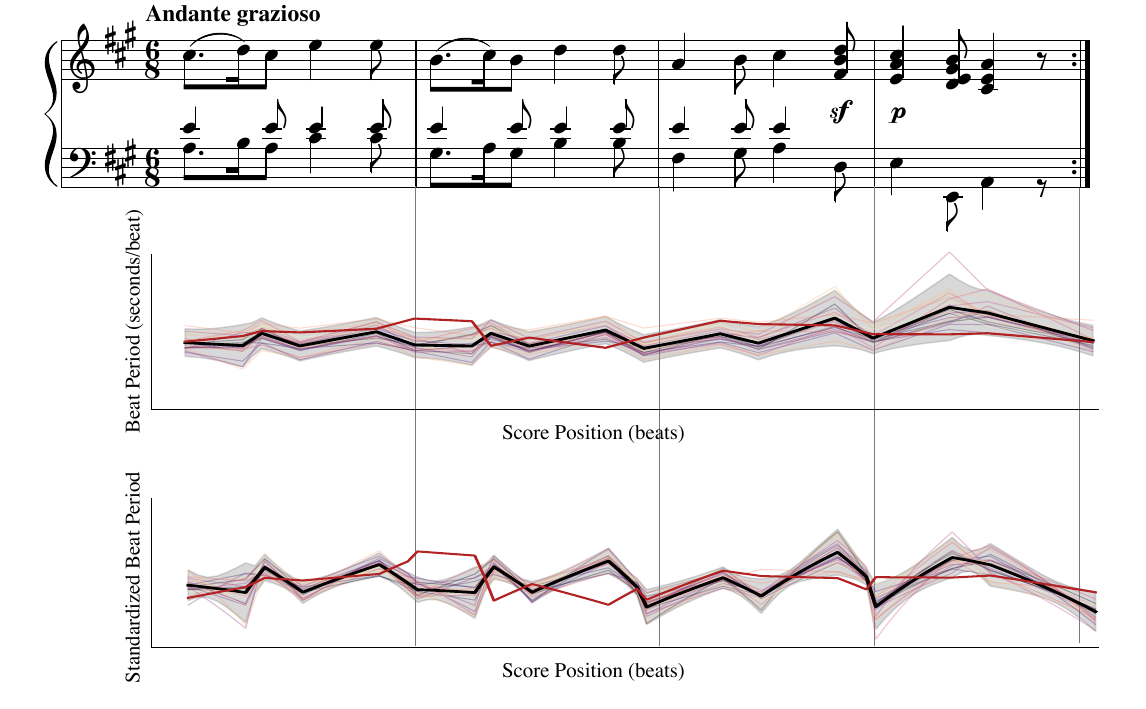}
\caption{At top: excerpt of Mozart's Piano Sonata K  331 Mv.~1 (Bars 5--8). Middle: non-standardized tempo curves. Bottom: mean-log standardized tempo curves (see Section \ref{sec:tce}). Colored lines represent tempo curves from the Vienna 4x22 dataset for the Mozart excerpt; black curves represent averaged tempo curves; red lines are randomly generated (non-musical) performances. The gray shaded area indicates one standard deviation above and below the average curve.}  
\label{fig:tempo_curves}
\end{center}
\end{figure*}

To aid the description of our experiments, we formalize reconstruction error-based evaluation as the following framework, as shown schematically in Figure \ref{fig:mod_eval}.
We consider a \emph{two-model evaluation framework} which asks the question: is performance P1 produced by 
Model1 "better" than P2 produced by Model2?
Concretely, the framework takes a triplet of two performances P1 and P2 of the same piece, one generated by each model, and computes their reconstruction error with respect to (wrt) a third expert RP of the same piece.

The standard evaluative argument of GMEPP is as follows: the model which produced the performance with smaller reconstruction error is favored and its performance is taken to be more musical.

This seemingly overly formal description of a simple and widely used evaluation technique allows us to formulate experiments \emph{about} evaluation by controlling specific elements. 
Specifically, we use the framework to evaluate \emph{models with controlled ground truth} wrt evaluation, i.e.,
with models which are known to better or worse.
Getting performances of good, musical models is straightforward, any human expert performance can be taken as such.
However, finding unmusical performances faces the performance research version of the Anna Karenina principle: all musical performance are (potentially) alike, but all unmusical performances are unmusical in their own way.
To mitigate the complexity, we opt for a type of randomization to create unmusical performances. 

Before we describe our process to create (and validate our choice of) unmusical performances in Section~\ref{subsec:randomization}, we briefly introduce the numerical representation of expressive performances in Section~\ref{subsec:num}.
We then connect our two main experiments to the framework in Section~\ref{subsec:exp1}.

\subsection{Numerical Representation of Performances}\label{subsec:num}

To capture nuances and deviation from the score in performances we use numerical features. 
Every performance yields sequences of measurements encoding an expressively relevant attribute, e.g., tempo.
The sequence contains values (e.g., the current beat period) for each note or score onsets (from now on broadly referred to as dimensions), i.e., performances can be different from others in each of these dimensions and distance metrics aggregate differences in each of these dimensions into a single value.
An example of the numerical sequence representation of performances in terms of beat period is illustrated in Figure~\ref{fig:tempo_curves}.
Performances differ from each other (vertically) in each dimension, i.e., at each score onset on the horizontal axis.

\begin{figure}
\includegraphics[width=1.0\linewidth]{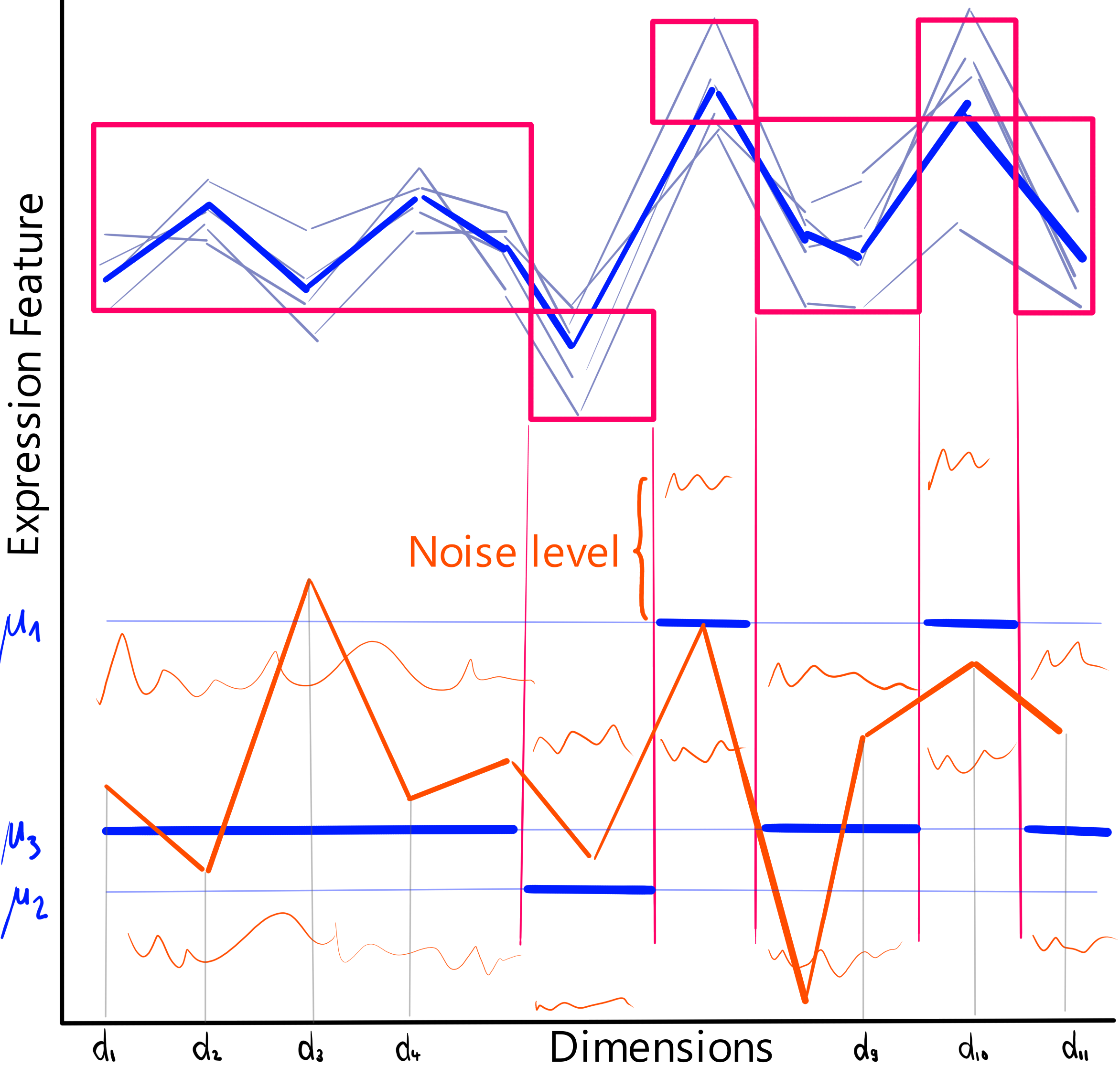}
\caption{
Illustration of the sampling process approximating the ball of expert performances with a mixture of three Guassian random variables. The average performance (opaque blue, top) is computed from expert performances (translucent blue, top) and segmented into quantiles (red boxes).
A randomized performance (orange, bottom) is then sampled from Gaussian distribution for each quantile, with a standard deviation controlled as noise level parameter.
}\label{fig:sampling}
\end{figure}

\subsection{Randomization within the Ball of Expert Performances}\label{subsec:randomization}

Given a number of expert performances of the same piece, anyone can be chosen as a reference performance.
This means that any other expert performance sits at some distance from the reference (in a high-dimensional space), some closer, some further away.
If we are able to synthesize performances with an expected distance no greater than this expert performance's distance from the reference, our performances would ---in expectation and according to the framework--- seem as good as an expert performance.
In other words, any performance in a (high-dimensional) ball around the average expert performance looks musical to this evaluation framework, given that
the ball diameter is the average distance between pairs of expert performances.

Our aim is thus to randomize performances that stay within this ball in expectation.
We approximate this using mixture of Gaussian random variables, set at the mean of quantiles of the average expert performance.
Figure~\ref{fig:sampling} illustrates this process from top to bottom.
We start by computing the average expression feature curve across performers of a chosen excerpt.
We then split the dimensions (horizontal axis) according to quantiles of the expression feature (vertical axis).
Finally, we define a Gaussian random variable for each quantile, defined by the mean of the expression curves within the quantile and a configurable standard deviation, which we refer to as noise level.
Setting the noise level allows for (probabilistic) control over the (expected) distance to an average expert performance.
Generally, the higher the noise level, the further the performance.

A possible result of randomization is shown in Figure \ref{fig:tempo_curves} using an excerpt of a Mozart Piano Sonata.
Expert performances are shown in gray and a generated randomized performance in red.
Note that the shaded area where many expert performances come to lie (mean performance and one standard deviation above and below) is \textit{not} an illustration of the high-dimensional ball defined by curves that do not exceed an average reconstruction error wrt the average performance.
Such high-dimensional balls are difficult to visualize, but intuitively large deviations in few dimensions are possible if the values in a majority of other dimensions fall very close to the reference.

Furthermore, note that this Gaussian mixture is not guaranteed to stay within this ball for general sequences or even any possible expression features sequences.
The performances of one excerpt might lie very closely, narrowing down the possibilities such that even noise level zero, i.e., a quantile-wise deadpan performance, is beyond the ball.
However, we never see this happen on our data.

\subsection{Experiments}\label{subsec:exp1}

In the first experiment, we are interested in the capacity of listeners to \textbf{discern slightly randomized performances} that look similar to expert performances to the quantitative framework.
Do listeners perceive these randomizations or are they too fine?
In a listening test, we present participants with several pairs of performance excerpts, each pair consisting of one expert and one randomized, and ask them to identify the expert performance among each pair.
The randomizations are created with precisely controlled error rates of the framework for each excerpt.

In the second experiment, we use the same randomized and expert performance pairs as in the previous one, however, with increased randomization strength and no excerpt-wise configuration of the randomization.
In this scenario, a listener should be overwhelmingly likely to identify the randomization, at the cost of the randomization also being more visible to the evaluation framework, i.e., the framework should identify more than 50 \% of human performances correctly.

This experiment addresses the second and third of our guiding questions: the \textbf{reliability}, i.e., the evaluative consistency, and the \textbf{validity}, i.e., the evaluative correctness, of the framework under various reference performances and and validity of the quantitative evaluation framework.


\section{Methods}\label{sec:tce}

The previous discussion of the framework remained abstract, in this sections we discuss concrete dataset, expression features, standardizations, metrics, and randomizations used in the two experiments.

\subsection{Datasets}\label{sec:datasets}

For our analysis we use excerpts of MIDI or MIDI-like recordings with performed notes matched to their corresponding score notes extracted from two datasets:

\noindent\textbf{Vienna 4x22:} This dataset was originally compiled by Goebl~\cite{vienna4x22}  and consists of 4 excerpts of solo piano pieces, each performed by 22 pianists. 
The excerpts are the first 21 bars of Chopin's Etude Op.~10 No.~3, the first 45 bars of Chopin's Ballade in F Op.~38, the first 36 bars (i.e., the theme) of Mozart's Piano Sonata in A K 331, and all 32 bars of Schubert D783 No.~15 (with repeats played). 
All performances were recorded on a B\"{o}sendorfer 290 SE Grand Piano as MIDI-like data and subsequently each played note matched to its respective score note.

\noindent\textbf{KAIST / International Piano-e-Competition:} This dataset consists of MIDI recordings of performances of several editions of the International Piano-e-Competition\footnote{formerly known as the Yamaha Piano-e-Competition~\url{http://piano-e-competition.com/default.asp}} for a number of which researchers at KAIST~\cite{Jeongetal:2019} collected and corrected scores in MusicXML format. 
All performances were recorded on Yamaha Disklavier instruments.
The scores and performances have been aligned by KAIST\footnote{\url{https://github.com/mac-marg-pianist/chopin_cleaned}} using Nakamura et al.'s HMM-based alignment tool~\cite{Nakamura:2017vy}.
We converted these alignments to Matchfile format~\citep{matchfile}, extracted the pieces for which more than eight - or more than five in the case of Bach's well-tempered clavier - performances exist and cleaned up the alignments. 

Taken together, this yields 33 pieces or excerpts thereof --- 16 by Fr\'ed\'eric Chopin, 8 by Johann Sebastian Bach, 5 by Ludwig Van Beethoven, 2 by Franz Liszt, 1 by Wolfgang Amadeus Mozart, and 1 by Franz Schubert --- with 40786 unique score onsets, each played at least 6 and 34 times for a total of 476 performances.

\subsection{Expression Features}

In order to compare expressive parameters, we do not work directly with note-wise onsets, dynamics etc, but we 
compute four expression features: two onset-wise features (tempo and velocity), and two note-wise features (timing and articulation).
These features are defined as follows:

\begin{itemize}
\item A \textbf{tempo} curve is derived by dividing the performed inter-onset interval (IOI) by the score IOI for every score onset, where the performed onsets are first averaged across note sharing a score onset.
Tempo curves encode a measure of the rate of change measured in seconds per beat (aka beat period).
\item Likewise, \textbf{dynamics} curves are computed as the average MIDI velocity of individual notes at each score onset.
\item We define \textbf{timing} as the note-wise deviation from an average onset time of notes at a common score onset (as used in the tempo computation above) in milliseconds. 
The timing of notes sharing a score onset sums thus to zero, the timing of notes unique at their onset is also zero.
\item We define \textbf{articulation} as the base-two logarithm of the played duration divided by the notated duration times the beat period.
\end{itemize}

These definitions are by no means universal, however, these or equivalent expression features are commonly used (see section 4.1 in \cite{CancinoChacon:2018po}).

\subsection{Standardization and Metric}

The literature provides many examples of standardization, factoring and smoothing of expressive parameter curves (e.g. \cite{Dixon:2006vi,li2015clustering, Chew:2012ty, tobudic03}).
Li et al.~\cite{li2015clustering} proposed a number of standardization techniques which they compared as parameters in a model selection test. 
We evaluate four standardization techniques: mean standardization, mean-log standardization, mean/variance standardization (aka sampling standard score), and no standardization, under mean squared error  (MSE). 
In the middle plot of Figure \ref{fig:tempo_curves} non-standardized tempo curves are shown, and in the bottom we show the same curves but mean-log standardized.
Note that the MSE of two series of data points - i.e. performances - $x_1$ and $x_2$ under mean variance standardization is equivalent to $2-2\times \rho(x_1,x_2)$, where $\rho$ is the Pearson correlation coefficient. 
The given test results hence also imply evaluation under correlation, another commonly used metric.

\subsection{Randomization}

For our experiments we use the following quantile and noise level settings.
The listening test uses quartiles ($\mathcal{Q}_1, \mathcal{Q}_2, \mathcal{Q}_3, \mathcal{Q}_4$), the noise level $\sigma$ is set for each excerpt individually to control the evaluative validity of the framework. 
Formally, the randomizations are sampled from:

\begin{equation*}
    x_{t} \sim
\left\{
	\begin{array}{ll}
		 \mathcal{N}(\mu_{1},\,\sigma^{2}) & \mbox{if } t \in  \{o \mid  y_o \in \mathcal{Q}_1\} \\
		 \mathcal{N}(\mu_{2},\,\sigma^{2}) & \mbox{if } t \in  \{o \mid  y_o \in \mathcal{Q}_2 \} \\
		 \mathcal{N}(\mu_{3},\,\sigma^{2}) & \mbox{if } t \in  \{o \mid  y_o \in \mathcal{Q}_3\} \\
		 \mathcal{N}(\mu_{4},\,\sigma^{2}) & \mbox{if } t \in  \{o \mid  y_o \in \mathcal{Q}_4\} \\
	\end{array}
\right.
\end{equation*}

The second experiment uses unequal quantiles (lowest 5\%, center 90\%, and highest 5\%), the noise level is set to the average standard deviation across performances $\overline{\sigma}$. 
Formally, the randomizations are sampled from:

\begin{equation*}
    x_{t} \sim
\left\{
	\begin{array}{ll}
		 \mathcal{N}(\mu_{1},\,\overline{\sigma}^{2}) & \mbox{if } t \in  \{o \mid P(y > y_o) \leq 0.05 \} \\
		 \mathcal{N}(\mu_{2},\,\overline{\sigma}^{2}) & \mbox{if } t \in  \{o \mid P(y < y_o) \leq 0.05 \} \\
		 \mathcal{N}(\mu_{3},\,\overline{\sigma}^{2}) & \mbox{else } 
	\end{array}
\right.
\end{equation*}

where the quantiles of dimensions $t$ are shown as sets constrained by the probabilities $P$ of curve values $y$ at these dimensions.


\begin{table*}
\begin{tabular}{|l|ll|ll|ll|ll|ll|ll|ll|ll|} \hline  
Feature & \multicolumn{4}{|c|}{Articulation}  & \multicolumn{4}{|c|}{Timing}  & \multicolumn{4}{|c|}{Tempo}  & \multicolumn{4}{|c|}{Velocity}  \\ \hline  
Noise level & \multicolumn{2}{|c|}{50}  & \multicolumn{2}{|c|}{90}  & \multicolumn{2}{|c|}{50}  & \multicolumn{2}{|c|}{90}  & \multicolumn{2}{|c|}{50}  & \multicolumn{2}{|c|}{90}  & \multicolumn{2}{|c|}{50}  & \multicolumn{2}{|c|}{90}  \\ \hline  
Number of answers & \multicolumn{2}{|c|}{240}  & \multicolumn{2}{|c|}{185}  & \multicolumn{2}{|c|}{215}  & \multicolumn{2}{|c|}{235}  & \multicolumn{2}{|c|}{215}  & \multicolumn{2}{|c|}{234}  & \multicolumn{2}{|c|}{238}  & \multicolumn{2}{|c|}{213}  \\ \hline  
Number correct & \multicolumn{2}{|c|}{148}  & \multicolumn{2}{|c|}{139}  & \multicolumn{2}{|c|}{126}  & \multicolumn{2}{|c|}{129}  & \multicolumn{2}{|c|}{147}  & \multicolumn{2}{|c|}{183}  & \multicolumn{2}{|c|}{125}  & \multicolumn{2}{|c|}{109}  \\ \hline  
Percentage correct & \multicolumn{2}{|c|}{61.67}  & \multicolumn{2}{|c|}{75.14}  & \multicolumn{2}{|c|}{58.60}  & \multicolumn{2}{|c|}{54.89}  & \multicolumn{2}{|c|}{68.37}  & \multicolumn{2}{|c|}{78.21}  & \multicolumn{2}{|c|}{52.52}  & \multicolumn{2}{|c|}{51.17}  \\ \hline  
Percentage probability & \multicolumn{2}{|c|}{0.01}  & \multicolumn{2}{|c|}{0.00}  & \multicolumn{2}{|c|}{0.22}  & \multicolumn{2}{|c|}{1.69}  & \multicolumn{2}{|c|}{0.00}  & \multicolumn{2}{|c|}{0.00}  & \multicolumn{2}{|c|}{3.82}  & \multicolumn{2}{|c|}{5.15}  \\ \hline  
Piece & \rot{Mozart k331}  & \rot{Schubert D783 no15}  & \rot{Mozart k331}  & \rot{Schubert D783 no15}  & \rot{Chopin op38}  & \rot{Schubert D783 no15}  & \rot{Chopin op38}  & \rot{Schubert D783 no15}  & \rot{Chopin op38}  & \rot{Schubert D783 no15}  & \rot{Chopin op38}  & \rot{Schubert D783 no15}  & \rot{Chopin op10 no3}  & \rot{Schubert D783 no15}  & \rot{Chopin op10 no3}  & \rot{Schubert D783 no15}  \\ \hline  
Excerpt start & \rotnothing{8}  & \rotnothing{0}  & \rotnothing{8}  & \rotnothing{0}  & \rotnothing{33}  & \rotnothing{17}  & \rotnothing{33}  & \rotnothing{17}  & \rotnothing{33}  & \rotnothing{0}  & \rotnothing{33}  & \rotnothing{0}  & \rotnothing{0}  & \rotnothing{17}  & \rotnothing{0}  & \rotnothing{17}  \\ \hline  
Excerpt duration & \rotnothing{8}  & \rotnothing{9}  & \rotnothing{8}  & \rotnothing{9}  & \rotnothing{12}  & \rotnothing{8}  & \rotnothing{12}  & \rotnothing{8}  & \rotnothing{13}  & \rotnothing{9}  & \rotnothing{13}  & \rotnothing{9}  & \rotnothing{6}  & \rotnothing{8}  & \rotnothing{6}  & \rotnothing{8}  \\ \hline  
Number of answers & \rotnothing{113}  & \rotnothing{127}  & \rotnothing{105}  & \rotnothing{80}  & \rotnothing{110}  & \rotnothing{105}  & \rotnothing{119}  & \rotnothing{116}  & \rotnothing{106}  & \rotnothing{109}  & \rotnothing{113}  & \rotnothing{121}  & \rotnothing{118}  & \rotnothing{120}  & \rotnothing{107}  & \rotnothing{106}  \\ \hline  
Number correct & \rotnothing{78}  & \rotnothing{70}  & \rotnothing{75}  & \rotnothing{64}  & \rotnothing{66}  & \rotnothing{60}  & \rotnothing{70}  & \rotnothing{59}  & \rotnothing{74}  & \rotnothing{73}  & \rotnothing{88}  & \rotnothing{95}  & \rotnothing{63}  & \rotnothing{62}  & \rotnothing{61}  & \rotnothing{48}  \\ \hline  
Percentage correct & \rotnothing{69.0}  & \rotnothing{55.1}  & \rotnothing{71.4}  & \rotnothing{80.0}  & \rotnothing{60.0}  & \rotnothing{57.1}  & \rotnothing{58.8}  & \rotnothing{50.9}  & \rotnothing{69.8}  & \rotnothing{67.0}  & \rotnothing{77.9}  & \rotnothing{78.5}  & \rotnothing{53.4}  & \rotnothing{51.7}  & \rotnothing{57.0}  & \rotnothing{45.3}  \\ \hline  
Percentage probability & \rotnothing{0.00}  & \rotnothing{3.65}  & \rotnothing{0.00}  & \rotnothing{0.00}  & \rotnothing{0.85}  & \rotnothing{2.68}  & \rotnothing{1.15}  & \rotnothing{7.27}  & \rotnothing{0.00}  & \rotnothing{0.01}  & \rotnothing{0.00}  & \rotnothing{0.00}  & \rotnothing{5.60}  & \rotnothing{6.80}  & \rotnothing{2.71}  & \rotnothing{4.84}  \\ \hline  
\end{tabular} 

\caption{Results of the listening test broken down hierarchically, with the top row identifying the four expression features.
The next five rows divide each feature into two noise levels and report from top to bottom: the noise level used, the total number of answers, the number of correct expert performer identification, the ratio of correct identification as percentage, and finally the probability (as percentage) of this outcome under the null hypothesis.
The next six rows report the same values again, albeit split down by excerpt. }  
\label{tab:results_lt}
\end{table*}

\section{Listener Discernment Experiment}\label{sec:experiments1}

Using a listening test we estimate the degree to which listeners are capable of discerning differences in performance expression features that look similar under the quantitative evaluation framework.

\subsection{Data}

For the listening test we extract excerpts of pieces of the Vienna 4x22 dataset.
We use two excerpts per expression feature, with all four expression features (tempo, timing, articulation, and velocity) being investigated, making for a total of eight excerpts.
The excerpts are chosen based on two considerations:
First, they need to cover enough musical material to be able to judge phrasing and timbre, but not be too long for the listeners. 
We opt for 8 - 10 measures.
Secondly, we extract all excerpts fulfilling the length criteria and measure their inter-performance correlation.
For each expression feature, we choose the excerpts with the highest and the lowest correlations, respectively.
For high correlation excerpts, performances are very consistent across performers, we thus expect the randomization ball to be small, and identification of randomized performances correspondingly harder.
We further double the number of test pairs by using two noise levels.
Noise level 50 refers to standard deviations set in the randomization such that the framework identifies 50 \% of the pairs correctly, i.e., the framework evaluates at chance level, the randomization is indistinguishable to the framework.
At noise level 90, the framework identifies 90 \% of randomizations.
We expect listeners to be able to identify the stronger randomizations (noise level 90) with greater ease.
Each of the eight excerpts is matched with 44 randomized performances, 22 at noise level 50, 22 at noise level 90.

\subsection{Listening Test}

Participants are provided with an online questionnaire of 16 pairs of performances, one for each test case, randomly sampled from the $22\times22$ possible (random $\times$ expert) pairings.
On the first page, listeners are instructed to the task --- listening to the two audio files and identifying the expert performance among them --- and presented the five items of the short Musical Training subsection of the Goldsmiths Music Sophistication self-assessment Index (GMSI).
Of the participants that completed the GMSI questions, $56\%$ engaged in regular practice of a musical instrument for 4 or more years and $69\%$ reported practicing their primary instrument for at least 2 hours per day.
Listeners can start, pause, stop, or rewind the audio excerpts at their leisure.
The possible answers include: performance 1 is the expert performance, performance 2 is the expert performance, and undecided.

\subsection{Results}

More than 250 listeners participate in the online study, with usable (unskipped) answers per (noise level $\times$ feature)-configuration ranging from 185 to 240.  
Table~\ref{tab:results_lt} presents the results of the listening test.
The table breaks down the answers hierarchically, with the top row identifying the four expression features studied.
The next five rows from the top divide each feature into two noise levels and report from top to bottom: the noise level used, the total number of answers, the number of correct expert performer identification, the ratio of correct identification as percentage, and finally the probability (as percentage) of this outcome under a binomial distribution with success probability of $0.5$, the distribution corresponding to the null hypothesis; listeners can't discern the expert performances.
The next six rows report the same values again, albeit further broken down by excerpt. 
For each excerpt we further note the starting point and duration in measures.

Most apparent from the results is that the inconsistency of listener discernment across features.
They largely fail to perform better than chance for timing and velocity, yet show clear (and statistically significant at p=0.01) discernment for articulation and tempo.
Furthermore, the noise level influences the results as assumend for tempo and articulation, but fails to influence the judgment of the other two in a significant way.
Addressing our first guiding question, listeners discerned randomization in both articulation and tempo which are indistinguishable under the evaluative framework (noise level 50).
However, the framework readily identifies stronger randomizations (noise level 90) in velocity and timing, which escape the listeners.

\section{Validity and Reliability Experiment}\label{sec:experiments2}

\begin{table*}
\resizebox{1.0\textwidth}{0.9\textheight} {
\rotb{
\begin{tabular}{|llrr|ccc|c|c|ccc|c|c|}
    \hline    
    \multicolumn{4}{|c|}{{\bfseries  global values }} &
    \multicolumn{5}{c|}{{\bfseries  values wrt dynamics curves }}& 
    \multicolumn{5}{c|}{{\bfseries  values wrt tempo curves }}\\ \hline
    \bfseries \rot{Piece} &  
    \bfseries \rot{Composer} &  
    \bfseries \rot{No e. perf.} &  
    \bfseries \rot{No Onsets} &  
    \bfseries \rot{$\overline{MSE}$ e. - e.} &  
    \bfseries \rot{$\overline{MSE}$ e. - r.} &  
    \bfseries \rot{$\overline{MSE}$ r. - r.} &  
    \bfseries \rot{reliability} & 
    \bfseries \rot{validity} &
    \bfseries \rot{$\overline{MSE}$ e. - e.} &  
    \bfseries \rot{$\overline{MSE}$ e. - r.} &  
    \bfseries \rot{$\overline{MSE}$ r. - r.} &  
    \bfseries \rot{reliability } & 
    \bfseries \rot{validity} 
    \\
    \hline
        Etude Op. 10 No 1 & Chopin & 26 & 1187 & 1.16 & 1.33 & 0.88 & 0.98 & 10.2     \%  & 1.11 & 1.19 & 0.75 & 0.94 & 30.6     \% \\ \hline
        Etude Op. 10 No 2 & Chopin & 11 & 729 & 1.23 & 1.39 & 0.95 & 0.58 & 16.6  \%  & 1.09 & 1.43 & 1.12 & 0.86 & 3.7  \% \\ \hline
        Etude Op. 10 No 4 & Chopin & 26 & 1145 & 0.96 & 1.24 & 0.74 & 0.94 & 1.8  \%  & 1.28 & 1.28 & 0.82 & 0.18 & 49.0  \% \\  \hline
        Etude Op. 10 No 5 & Chopin & 11 & 894 & 0.96 & 1.22 & 0.67 & 0.87 & 3.6  \%  & 0.94 & 1.03 & 0.48 & 0.32 & 28.6  \% \\ \hline
        Etude Op. 10 No 8 & Chopin & 28 & 1380 & 1.03 & 1.28 & 0.78 & 0.9 & 3.1  \%  & 1.14 & 1.25 & 0.86 & 0.49 & 22.6  \%  \\ \hline
        Etude Op. 10 No 12 & Chopin & 13 & 1264 & 0.76 & 1.04 & 0.42 & 0.98 & 0.5  \%  & 0.85 & 0.88 & 0.28 & 0.2 & 36.9  \%  \\ \hline
        Etude Op. 25 No 10 & Chopin & 12 & 910 & 0.39 & 1.06 & 0.28 & 1.0 & 0.0  \%  & 0.45 & 0.81 & 0.2 & 1.0 & 0.0  \%  \\ \hline
        Etude Op. 25 No 11 & Chopin & 25 & 1878 & 0.94 & 1.17 & 0.69 & 0.86 & 4.3  \%  & 1.13 & 1.33 & 0.99 & 0.72 & 8.9  \% \\ \hline
        Barcarolle Op. 60 & Chopin & 9 & 1597 & 0.53 & 1.06 & 0.35 & 1.0 & 0.0  \%  & 0.66 & 0.77 & 0.16 & 0.45 & 18.6  \%  \\ \hline
        Sonata No. 3 Op. 58 2nd & Chopin & 9 & 879 & 0.75 & 1.03 & 0.42 & 0.99 & 0.2  \% & 0.54 & 0.87 & 0.25 & 1.0 & 0.0  \%  \\ \hline
        Sonata No. 3 Op. 58 4th & Chopin & 9 & 2506 & 0.92 & 1.17 & 0.55 & 1.0 & 0.0  \%  & 1.03 & 1.12 & 0.62 & 0.3 & 22.6  \% \\ \hline
        Scherzo No. 2 Op. 31 & Chopin & 12 & 2874 & 0.49 & 1.05 & 0.35 & 1.0 & 0.0  \%  & 0.57 & 0.79 & 0.22 & 0.97 & 0.8  \%  \\ \hline
        Ballade No 1 Op. 23 & Chopin & 18 & 2174 & 0.49 & 0.97 & 0.31 & 1.0 & 0.0  \%  & 0.35 & 0.66 & 0.11 & 1.0 & 0.0  \% \\ \hline
        Ballade No 4 Op. 52 & Chopin & 12 & 2583 & 0.45 & 1.03 & 0.31 & 1.0 & 0.0  \%  & 0.51 & 0.83 & 0.21 & 1.0 & 0.0  \%  \\ \hline
        WTC BWV 848 F. & Bach & 9 & 831 & 1.15 & 1.36 & 0.84 & 0.47 & 15.4  \%  & 0.99 & 1.37 & 1.04 & 1.0 & 0.0  \%  \\ \hline
        WTC BWV 857 F. & Bach & 6 & 753 & 0.93 & 1.17 & 0.56 & 0.79 & 5.6  \%  & 0.8 & 1.11 & 0.44 & 0.36 & 20.3  \%  \\ \hline
        WTC BWV 860 F. & Bach & 6 & 988 & 1.03 & 1.29 & 0.82 & 0.78 & 7.4  \%  & 0.53 & 1.09 & 0.55 & 1.0 & 0.0  \%  \\ \hline
        WTC BWV 889 F. & Bach & 6 & 567 & 1.34 & 1.4 & 1.02 & \bfseries \textcolor{rel}{0.1} & \bfseries \textcolor{val1}{39.9  \%} & 0.79 & 1.27 & 0.83 & 1.0 & 0.1  \%  \\ \hline
        WTC BWV 848 P. & Bach & 9 & 610 & 1.04 & 1.24 & 0.67 & 0.57 & 14.6  \%  & 1.01 & 1.52 & 1.3 & 0.99 & 0.1  \%  \\ \hline
        WTC BWV 857 P. & Bach & 6 & 317 & 1.04 & 1.24 & 0.66 & 0.46 & 16.6  \%  & 0.68 & 1.15 & 0.84 & 1.0 & 0.0  \%  \\ \hline
        WTC BWV 860 P. & Bach & 6 & 427 & 0.96 & 1.17 & 0.6 & 0.65 & 9.9  \%  & 0.99 & 1.37 & 0.86 & 0.83 & 5.8  \%  \\ \hline
        WTC BWV 889 P. & Bach & 6 & 595 & 0.88 & 1.21 & 0.81 & 1.0 & 0.0  \%  & 0.98 & 1.15 & 0.81 & 0.47 & 15.5  \%  \\ \hline
        T. Etude S. 139 No 10 & Liszt & 16 & 1842 & 0.66 & 1.09 & 0.44 & 1.0 & 0.0  \%  & 0.64 & 0.92 & 0.31 & 0.99 & 0.3  \%  \\ \hline
        P. Etude S.141 No 1 & Liszt & 34 & 439 & 1.19 & 1.36 & 0.94 &  \bfseries \textcolor{rel}{0.47} & 21.2  \%  & 1.37 & 1.58 & 1.35 &  \bfseries \textcolor{rel}{0.3} & 29.3  \% \\ \hline
        Sonata No 3 Op. 2/3 1st & Beethoven & 9 & 2190 & 0.45 & 0.99 & 0.27 & 1.0 & 0.0  \%  & 0.46 & 1.09 & 0.37 & 1.0 & 0.0  \%  \\ \hline
        Sonata No 4 Op. 7 1st & Beethoven & 9 & 2046 & 0.76 & 1.06 & 0.42 & 1.0 & 0.0  \% & 1.22 & 1.19 & 0.65 &  \bfseries \textcolor{rel}{0.13} & 61.1  \%  \\ \hline
        Sonata No 17 Op. 31/2 1st & Beethoven & 11 & 1376 & 0.6 & 1.02 & 0.38 & 1.0 & 0.0  \%  & 0.18 & 0.38 & 0.04 & 0.99 & 0.4  \%\\ \hline
        Sonata No 18 Op. 31/3 1st & Beethoven & 9 & 1707 & 0.64 & 1.0 & 0.38 & 1.0 & 0.0  \%  & 1.0 & 0.81 & 0.17 & 0.6 & \bfseries \textcolor{val1}{88.3  \%} \\ \hline
        Sonata No 21 Op. 53 1st & Beethoven & 25 & 3452 & 0.5 & 1.09 & 0.4 & 1.0 & 0.0  \%  & 0.75 & 1.13 & 0.56 & 0.82 & 7.0  \% \\ \hline
        Etude Op. 10 No. 3 & Chopin & 22 & 162 & 0.34 & 0.97 & 0.29 & 1.0 & 0.0  \%  & 0.43 & 0.83 & 0.44 & 0.97 & 0.8  \%  \\ \hline
        Ballade No 2 Op. 38 & Chopin & 22 & 197 & 0.64 & 1.05 & 0.57 & 0.99 & 0.3  \%  & 0.14 & 0.86 & 0.27 & 0.99 & 0.3  \%  \\ \hline
        Sonata No. 11 K. 331 & Mozart & 22 & 178 & 0.66 & 1.14 & 0.6 & 0.96 & 1.0  \%  & 0.47 & 0.9 & 0.43 & 0.92 & 2.2  \%  \\ \hline
        G. Dance D. 783 No. 15 & Schubert & 22 & 109 & 0.6 & 1.19 & 0.56 & 1.0 & 0.1  \% & 0.66 & 1.14 & 0.65 & 0.79 & 6.8  \%  \\ \hline
        \bfseries Dataset &  & \bfseries 476 & \bfseries 40786 & \bfseries 0.8 & \bfseries 1.15 & \bfseries 0.57 &\bfseries  \textcolor{rel}{0.85} & \bfseries \textcolor{val1}{5.2     \%}  & \bfseries 0.78 &\bfseries 1.06 &\bfseries  0.58 & \bfseries \textcolor{rel}{0.73} & \bfseries \textcolor{val1}{14.0     \%} \\ \hline

\end{tabular}
}
}

\caption{The results of the validity and reliability tests by piece. The abbreviations used are: e. = expert; perf. = perfomance; r. = random; corr. = Pearson correlation coefficient; mod. = model evaluation; No = number; Op. = Opus; F. = Fugue; P. = Prelude; WTC = Well-Tempered Clavier; BWV = Bach-Werke-Verzeichnis; G. = German; D. = Deutsch catalogue; K. = Köchelverzeichnis; S. = Searle catalogue}  
\label{tab:results}
\end{table*}

This experiment addresses our guiding questions two and three, concerning the reliability and validity of reconstruction error-based evaluations under different reference performances, respectively.
All experiments are carried out with respect to two of the performances' expressive parameters, namely onset-wise tempo and dynamics curves.

See Figure \ref{fig:mod_eval} for a schematic representation of the frameworks.
We use this framework to evaluate expert performances against randomized ones.
For each piece in the combined datasets described above (see Section \ref{sec:datasets}), we create 64 randomized ones.
The randomization starts from the average expert perfromance and follows the process described in section~\ref{subsec:randomization} and used in the listening test, albeit with one major difference:
the randomization follows a mixture of three Gaussians corresponding to the top 5 \%, bottom 5 \%, and center 90 \% quantiles, the noise level is set to the overall average standard deviation of the expression features for each piece.
Given the results of the listening test, we assume a listener to be overwhelmingly likely to identify the randomization, at least for the tempo curves.

\subsection{Reliability and Validity}

Using the described ground truth models, we compute validity and reliability values for the given evaluation frameworks.
We define \textbf{reliability} as the consistency of the evaluation framework under changes of references and across a variety of pieces, independent of the correctness of this result.
Given a human expert performance (produced by the 'musical' model) and a random sequence (produced by the 'unmusical' model), does the framework consistently favor the same model wrt different reference performances?
This consistency is quantified as average correlation of the binary output of the two-model evaluation (0 = model 1 has smaller MSE,1 = model 2 has smaller MSE) wrt different targets.
This is interpretable as inter-reference-performance correlation of evaluation framework. 
A perfectly reliable evaluation always favors the same model independent of RP.

We define the \textbf{validity} of the frameworks as the extent to which they accurately recover the ground truth.
A perfectly valid evaluation will always favor the expert performance and reject the randomized sequence.
Numerically, validity is estimated by the ratio of tests that erroneously recover the randomized performance over all possible reference, test and random performance combinations.
As for reliability, we compute and compare this number across a variety of pieces.

All in all, then, both tempo and dynamics under four standardizations are evaluated in two tests over 33 pieces. 
This amounts to a total of $2\times4\times2\times33=792$ experiments. Every test is carried out for the $n$ reference performances, $n-1$ test expert performances, and 64 randomly sampled performances, where n is the number of expert performances available for the respective piece.

\subsection{Results}\label{sec:results}

In this section, we present the results of selected tests. 
Results are reported in Table \ref{tab:results}, one part for dynamics curves, and another for tempo curves.
MSE between expression features under mean variance standardization, i.e. the standard score per performance excerpt, proved most beneficial for the framework's discernment capacity and is hence used throughout the experiments.
Each row in Table \ref{tab:results} represents a piece, the values given in the first four columns are as follows.
The name and opus number of the piece and its composer.
The number of expert performances, the number of their shared onsets.

The following four columns are given once for tempo curves and once for dynamics curves.
The mean of three MSE distributions: 
the inter expert performance MSEs, the MSEs between expert performances and randomized performances, and the MSEs among randomized performances.
The next column reports the reliability of the two-model evaluation as the mean of correlations among the two model tests over different target performances.
Lastly, the validity of the framework is given as the percentage of randomly sampled performances with lower MSE than a given expert performance.

\subsubsection{Reliability}
Values in Table \ref{tab:results} relating to important aspects of reliability are colored in \textbf{\textcolor{rel}{red}}.
The average correlation of all two-model evaluations wrt dynamics curves is 0.85, the highest value being 1.0 and the lowest 0.09 (Table \ref{tab:results}, col. 8).
The average correlation of all two-model evaluations wrt tempo curves is 0.73, the highest value being 1.0 and the lowest 0.13 (col. 14).
Generally, there is agreement in a majority of pieces and less reliability in a minority. 
For 13 pieces, the correlation of evaluations drops below 0.5 wrt tempo or dynamics, highlighting high variation across pieces.

The pieces exhibiting low reliability differ between tempo curve and dynamics curves tests.
Only one piece (Grande Etude de Paganini S.141 No 1) shows correlation below 0.5 in both tempo and dynamics curves.

\subsubsection{Validity}
Important values in Table \ref{tab:results} relating to the validity of the two-model evaluation are colored in \textbf{\textcolor{val1}{blue}}.
The two-model evaluation validity tests show an average of 5.3 \% of comparisons wrt dynamics, and an average of 14.0 \% of comparisons wrt tempo, favoring the randomly sampled performance (col. 9/14).
The average for all pieces is not weighted by the number of expert performances or tests.
The percentages vary greatly from perfect recovery of all expert performances to 88.3 \% of evaluations favoring random performances.
Again, valid evaluation wrt tempo does not imply valid evaluation wrt dynamics and vice versa.
15 pieces exhibit good performance of the framework with rejection of random performances in more than 90 \% of cases for both tempo and dynamics.


\section{Discussion and Conclusions}\label{sec:discussion}

Performance data is complex and sometimes more opaque than apparent at first glance.
Not without reason have researchers interested in performance practice and computational performance modelling spent decades dissecting the minutiae of phrasing, melody lead, pedalling, to name 
just a few aspects.
To better appreciate the breadth of issues, we briefly discuss several research directions.

Directly implied in our investigations are computational models of expressive performance, we refer to \cite{CancinoChacon:2018po} and \cite{lerch_music_2019} for a comprehensive overview.
For an overview of methods for evaluating computational models of expressive performance, we refer the reader to~\cite{Bresin:2013ty}.

Other cues come from performance research related to listener judgments, e.g., the seminal work by Repp~\cite{Repp1997} which presents evidence suggesting that listeners prefer average performances.
Wesolowski et al.~\cite{Wesolowski:2016kk} present a critical view of listeners' aesthetic judgments as a methodological tool for evaluating the differences in Jazz ensemble performances by analyzing their ratings' variability.
The music psychology literature provides evidence showing that the assessment of the (aesthetic) quality of a performance depends not only on the auditory component of a performance (e.g., \cite{Platz:2012dq}). 

Performance practice research is also interested in an entirely different type of perceptual classification of performance, namely semantic descriptor of expressive performance, or, more commonly, instrumental timbre.
By means of example, we refer to the sequence of studies undertaken by Bernays et al.~\cite{bernays2010expression,bernays2013expressive,bernays2014investigating}, or more recently and from within the MIR community~\cite{cancino2020characterization}.
Besides verbal descriptions, quantitative performance research often takes the form of detailed analyses of expression features in specific contexts. Exemplary work was carried out by Goebl et al.~\cite{Goebl2009, vienna4x22}, e.g., their work on the sources of melody lead~\cite{goeblLead}.

From a music education perspective, Gururani et al.\cite{gururani_analysis_2018} investigate quantitative descriptors for assessing the quality of performance.
Pati et al. \cite{pati_assessment_2018} present a deep learning based approach to assess student music performance.

Our tests add some bits to the knowledge surrounding measured expressive performances and their generative models.
They indicate that MSE based model evaluation is not necessarily reliably favoring the same performance wrt different targets.
Furthermore, MSE based model evaluation is not dependably capable of discerning expert performances from randomized performances.
The pieces under examination show great variability both wrt to the  tests, as well as wrt closeness of expert performances.
Listeners perceive randomizations in articulation and tempo that escape the evaluation framework, but they do not notice randomizations in velocity and microtiming with the same acuity.
Reasons for this can be sought both in the perception as well as in the production of expressive performances.

How then can automatic, quantitative evaluation be improved?
Our experiments and experience allow only for tentative answers, but answers they still are:
Most settings seem to benefit from more fine-grained evaluations.
Shorter excerpts tend to give more reliable and valid results and are better suited to localize errors.
If multiple performances are available,
test excerpts can be chosen which have high internal consistency, i.e., high inter-performer correlation or low inter-performer MSE, respectively.
Ideally these excerpts can relate to specific and discussed performance issues like phrasing, clear voices, specific timbre, etc.
Formulated in the negative, researchers should avoid resting their evaluative arguments on aggregated absolute errors across large, undocumented test dataset splits.
Such numbers carry too little information about the models under scrutiny.

Even better evaluation could plausibly be achievable with \textit{distributional metrics},
e.g.~the probability of generated performances under a Gaussian process (GP) regressor fitted with expert performances or inversely the likelihood of a generative GP model, like the model Teramura et al.\cite{Teramura:2008tz} proposed, for test performances. 
In a similar vein, trained neural network (NN) discriminators seem a promising avenue for future research.
However, neither tractable (GP) nor untractable (NN) approaches are a priori connected to listener judgment.

This is by no means an exhaustive discussion of issues surrounding the perception, characterization, and quantification of expressive performance, but we hope it serves to gain an appreciation of the intricacies of this data.
Prospective as well as seasoned researchers in the field of GMEPP do well in reminding themselves of these facts:
piano performance are aesthetically, culturally and axiologically rich, dynamic, and complex musical objects. 

\begin{acks}
This work is supported by the European Research Council (ERC) under the EU’s Horizon 2020 research \& innovation programme, grant agreement No.~101019375 (“Whither Music?”), and the Federal State of Upper Austria (LIT AI Lab).
\end{acks}

\bibliographystyle{ACM-Reference-Format}
\bibliography{cc}

\end{document}